\documentclass[11pt,twoside]{article}
\usepackage{asp2010}

\resetcounters

\begin{document}

\title{Determining the Metallicity of Low-Mass Stars and Brown Dwarfs: Tools for Probing Fundamental Stellar Astrophysics, Tracing Chemical Evolution of the Milky Way and Identifying the Hosts of Extrasolar Planets}
\author{Andrew A. West$^{1,10}$, John J. Bochanski$^{2,3}$, Brendan P. Bowler$^4$, Aaron Dotter$^5$, John A. Johnson$^6$, Sebastian L\'epine$^7$, B{\'a}rbara Rojas-Ayala$^{8}$, Andreas Schweitzer$^9$
\affil{$^1$Department of Astronomy, Boston University, 725 Commonwealth Avenue, Boston, MA 02215, USA, email: aawest@bu.edu}
\affil{$^2$Astronomy \& Astrophysics Dept., Pennsylvania State
  University, 525 Davey Lab, University Park, PA, 16802, USA, email: jjb29@psu.edu}
\affil{$^3$Kavli Institute for Astrophysics and Space Research,
  Massachusetts Institute of Technology, Building 37, 77 Massachusetts
  Avenue, Cambridge, MA 02139, USA}
\affil{$^4$Institute for Astronomy, University of Hawai`i; 2680 Woodlawn Drive, Honolulu, HI 96822, USA, email: bpbowler@ifa.hawaii.edu}
\affil{$^5$Space Telescope Science Institute,3700 San Martin Dr., Baltimore, MD, 21218, USA, email: aaron.dotter@gmail.com}
\affil{$^6$Department of Astrophysics, California Institute of Technology, MC 249-17, Pasadena, CA 91125, USA, email: johnjohn@astro.caltech.edu}
\affil{$^7$Department of Astrophysics, Division of Physical Sciences, American Museum of Natural History, Central Park West at 79th Street, New York, NY 10024, USA, email: lepine@amnh.org}
\affil{$^8$Department of Astronomy, Cornell University, 610 Space Sciences Building, Ithaca, NY 14853, USA, email: babs@astro.cornell.edu}
\affil{$^9$Hamburger Sternwarte, University of Hamburg, Gojenbergsweg 112, D-21029, Hamburg, Germany, email: Andreas.Schweitzer@hs.uni-hamburg.de}
\affil{$^{10}$Visiting Investigator, Department of Terrestrial Magnetism, Carnegie Institute of Washington, 5241 Broad Branch Road, NW, Washington, DC 20015, USA}
}

\begin{abstract}
  We present a brief overview of a splinter session on determining the
  metallicity of low--mass dwarfs that was organized as part of the
  Cool Stars 16 conference.  We review contemporary spectroscopic and
  photometric techniques for estimating metallicity in low--mass dwarfs
  and discuss the importance of measuring accurate metallicities for
  studies of Galactic and chemical evolution using subdwarfs,
  creating metallicity benchmarks for brown dwarfs, and searching for
  extrasolar planets that are orbiting around low--mass dwarfs.  In
  addition, we present the current understanding of the effects of metallicity on
  stellar evolution and atmosphere models and discuss some of the
  limitations that are important to consider when comparing theoretical
  models to data.
\end{abstract}

\section{Introduction}

Low--mass dwarfs are the most numerous stellar constituents of the
Milky Way and have main sequence lifetimes that exceed the current age
of the Universe (at least for those that are not brown dwarfs). They
therefore form an important laboratory for probing the structure
and evolution of the Milky Way's disks. Because of their ubiquity,
cool dwarfs may represent the largest population of stars with
orbiting planets, especially low--mass planets in their respective
habitable zones, which are considerably closer for cool dwarf
systems. In addition, the diminutive sizes of these stars makes the
detection of transiting planets easier than for higher mass stars (for
any given planetary radius).  Previous results have demonstrated that
planets are more likely to be found orbiting metal-rich stars
\citep[e.g.][]{fischer05}. There were preliminary indications that the
M dwarfs with known planets had sub-solar metallicities
\citep{2005A&A...442..635B,bean06}, in stark contrast to their high-mass
counterparts. However, recent results have shown that the M dwarfs
with attending planets appear to be metal-rich
\citep[see Section 2; ][]{2009ApJ...699..933J}. With low--mass dwarfs becoming important
sites for planet hunting
\citep[e.g. MEarth;][]{irwin09,endl03,johnson07}, the observational
efficiency of these searches could be vastly increased with prior
knowledge of stellar metallicity.

Because the ages of low--mass dwarfs span the lifetime of the Milky
Way, they can provide important insight into the history and evolution
of the Galaxy. Recent improvements in kinematic modeling and magnetic
activity analysis have provided enhanced statistical age estimates for
populations of low--mass dwarfs \citep{west06,west08}. Coupled with
metallicity information, these ages can provide valuable insight into
the chemical evolution history of the Milky Way disks. Without large
samples of low--mass dwarfs, the utility of the statistically derived
ages is limited. Fortunately, the advent of large surveys such as SDSS and
2MASS has produced photometric samples of low--mass dwarfs that number
in the tens of millions \citep{boo10} and spectroscopic samples that
contain more than 70,000 M dwarfs \citep{west08, kruse10, west10} and
almost 500 L dwarfs \citep{schmidt_sam}. In addition, these large
catalogs of low--mass dwarfs have identified significant samples of
metal--poor subdwarfs. The detailed metallicities of these
objects, coupled with their kinematic distributions, establish
important constraints on the structure and composition of the Milky
Way halo.

Historically, the metallicity of low--mass dwarfs has been an elusive
fundamental property due to the complex atmospheres of M, L and T
dwarfs that have restricted the accuracy of detailed model
atmospheres. Over the past several years, new observational techniques
as well as independent theoretical advancements in atmospheric models
have produced results that appear to link the metallicity of low--mass
dwarfs to both their photometric and spectroscopic properties
\citep[e.g.,][]{bean06,2005A&A...442..635B, woolf06,2009ApJ...699..933J,
  phoenix10,2010ApJ...720L.113R}. While these relations and the
resulting metallicities provide fundamental measurements for stellar
astrophysics, they also play a crucial role in studies of Galactic
evolution and the environments that host extrasolar planets.

\section{Calibrating M Dwarf Metallicity using Photometry}

Several previous studies have estimated M dwarf metallicities using
wide binary pairs that consist of both an M dwarf and a higher mass
star \citep[e.g.,][]{bean06,2005A&A...442..635B, woolf06}.  Because
binaries are assumed to be both coeval and have the same metallicity,
the composition of the higher mass star (which can be accurately
derived from comparison to theoretical models) can be applied to the
companion M dwarf.  Some of these studies have used optical and
infrared spectroscopy to tie spectroscopic features to a metallicity
scale \citep[e.g.,][]{bean06, woolf09,
  2010ApJ...720L.113R}.  Although the spectroscopic method has shown
great promise in deriving M dwarf metallicities (see Section 3),
spectroscopy is considerably more time consuming than photometry and
may not be easy to obtain for large samples of stars.

\begin{figure}
 \plotone[width=0.8\textwidth]{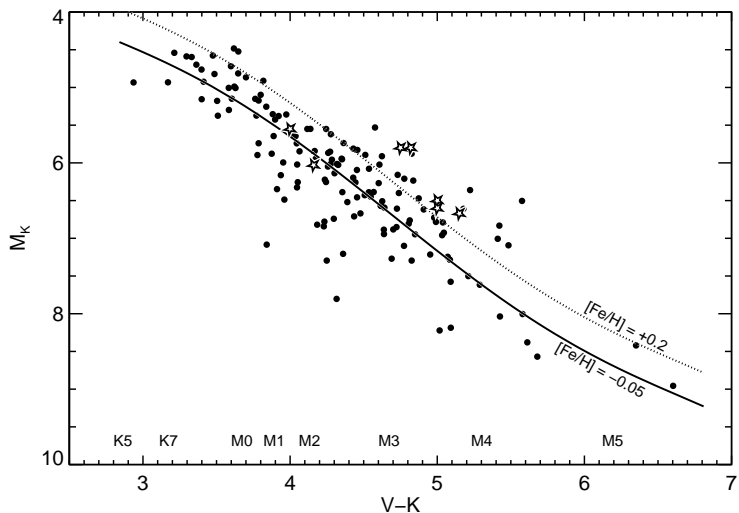}
 \caption{The low--mass target stars of the California Planet Survey in the 
$M_K$ vs $V-K$ plane (filled circles), and the M dwarfs known to
harbor one or more gas giant planets (five-point stars). The 
isometallicity contours for [Fe/H] = 0 (solid) and [Fe/H] = +0.2 
(dotted line) are based on the broad-band photometric metallicity
calibration of \citet{2009ApJ...699..933J}.}
 \label{johnson}
\end{figure}

\citet{2005A&A...442..635B} used M dwarfs in wide binaries to derive a
relation between the absolute $K$-band magnitude and the $V-K$ color
(higher metallicity M dwarfs are slightly brighter at a given color).
Given the large number of M dwarfs for which there exist photometric
observations, this relation may prove exceedingly useful.  However,
there were 2 problems with resulting analyses: 1) using the
\citet{2005A&A...442..635B} relation, planet hosting M dwarfs appeared
to be metal poor compared to their FGK star counterparts; and 2) and
the relation yielded a mean metallicity of M dwarfs in the solar
neighborhood that was almost 0.1 dex below the mean [Fe/H] of higher
mass stars.  These discrepancies were resolved by
\citet{2009ApJ...699..933J}, who discovered a systematic uncertainty
in the photometry used by \citet{2005A&A...442..635B}.
\citet{2009ApJ...699..933J} used corrected photometry to re-derive a
relation between the metallicity, $M_K$ and $V-K$ color of M dwarfs
\citep[see also][]{2010A&A...519A.105S}.

Figure \ref{johnson} shows the low--mass target stars of the California
Planet Survey in the $M_K$ vs $V-K$ plane (filled circles), and the M
dwarfs known to harbor one or more gas giant planets (five-point
stars). The isometallicity contours for [Fe/H] = 0 (solid) and [Fe/H]
= +0.2 (dotted line) are based on the broad-band photometric
metallicity calibration of \citet{2009ApJ...699..933J}. The
distribution of stars illustrates the tendency of planet-hosting M
dwarfs to be metal-rich compared to stars in the Solar
Neighborhood. The planet-metallicity relationship therefore holds for
M dwarfs as well as Sun-like FGK stars.

\begin{figure}
\plotone[width=0.8\textwidth]{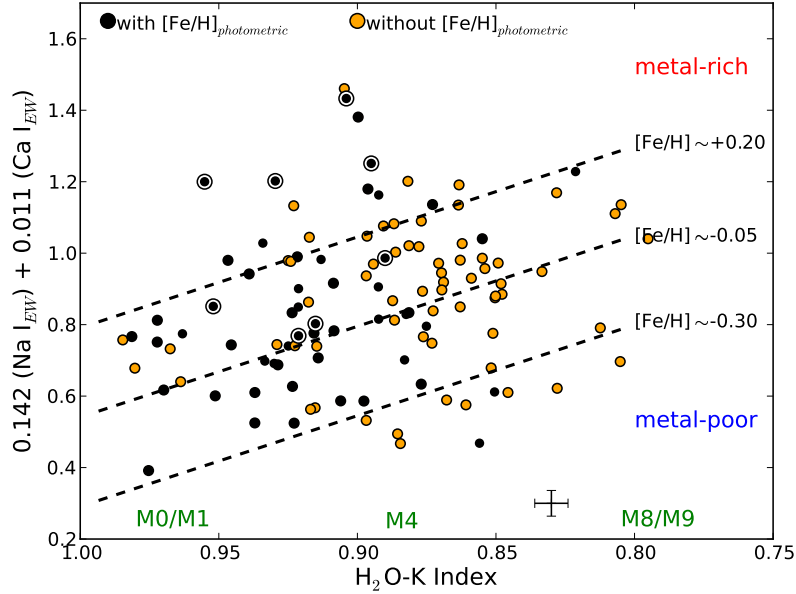}
\caption{A linear combination of the EWs of the Ca {\footnotesize I}
  and Na {\footnotesize I} features versus the H$_{2}$O-K index for
  northern 8 pc M-dwarfs. The black dots represent M dwarfs with
  photometric metallicities and the yellow dots represent M dwarfs
  with only near-infrared (NIR) spectroscopic metallicities. The big black dots (with circles)
  represent the M dwarf planet hosts. Typical errors in EWs and
  H$_{2}$O-K index are represented by the error bars. The dashed lines
  in the top panel are iso-metallicity contours for [Fe/H] values of
  -0.30, -0.05 and +0.20, calculated from the NIR [Fe/H]
  calibration. The NIR [Fe/H] calibration allows to cover a larger
  sample of cooler and distant M dwarfs (yellow dots)}
\label{babs}
\end{figure}

\section{Calibrating M Dwarf Metallicity using Infrared Spectroscopy}

Most of the attempts to estimate the overall metal content of M dwarfs
have been performed at visible wavelengths
\citep[e.g.,][]{1997AJ....113..806G, 2005A&A...442..635B,
  2009ApJ...699..933J}. Since M dwarfs are optically faint, this
limited past analyses to early-type M dwarfs and few specific nearby
stars, which are bright and have accurate parallaxes. To avoid this
limitation, \citet{2010ApJ...720L.113R} developed a near--infrared (NIR) [Fe/H]
spectroscopic calibration using strong absorption features in the
$K$-band spectra of M dwarfs. \citet{2010ApJ...720L.113R} adopted a
similar approach to that of \citet{2005A&A...442..635B} and
\citet{2009ApJ...699..933J}, assuming that binary systems share the
same metallicity since both components formed from the same original
molecular cloud. Seventeen FGK$+$M binary systems in the SPOCS catalog
\citep{2005ApJS..159..141V} were used as metallicity calibrators. The
NIR [Fe/H] calibration uses the Equivalent Widths (EWs) of the Na
{\footnotesize I} doublet and the Ca {\footnotesize I} triplet, and a
water absorption index \citep[H$_2$O,][]{2010ApJ...722..971C} to
differentiate between metal-rich and metal-poor M dwarfs
($\sigma$$\sim$0.15 dex). The results obtained with the NIR
spectroscopic [Fe/H] calibration are in agreement with the results
obtained with the photometric calibration by
\citet{2009ApJ...699..933J}. The eight M dwarf planet hosts analyzed
by \citet{2010ApJ...720L.113R} have metallicities higher than -0.05
dex, with the Jovian planets hosts being more metal-rich that their
Neptune analogs. This corroborates the \citet{2009ApJ...699..933J}
conclusion that planets are found preferentially around metal-rich
stars, like in their Sun-like counterparts.


As a moderate resolution $K$-band spectrum can be efficiently obtained
for most M dwarfs with current spectrographs (e.g. TripleSpec, FIRE),
the NIR [Fe/H] calibration allows observations of
cooler and distant M dwarfs (Figure \ref{babs}). Thus, this technique will
enable the identification of likely planet hosts at lower masses than
is possible with optical [Fe/H] techniques. However, the NIR [Fe/H]
calibration is currently limited to M dwarf spectral types earlier than $\sim$ M7
and [Fe/H] $>$ -0.7 , due to the lack of FGK dwarf$/$late-type M dwarf wide
binary systems with measured spectroscopic metallicities, and
subdwarfs with $\lambda/\Delta\lambda$$\approx$3000 NIR spectra, to be
used as calibrators.

\section{Spectral Features of Low-Metallicity Brown Dwarfs}

Brown dwarfs are expected to have a similar metallicity distribution
to the stellar components of our Galaxy, but reliably determining the
chemical compositions of individual brown dwarfs is a difficult task.
Atmospheric models remain largely untested at non-solar metallicities,
and there are no known benchmark brown dwarf companions to stars with
significantly super- or sub-solar chemical compositions ([Fe/H]
$\gtrsim$ +0.3 or [Fe/H] $\lesssim$ --0.3).  The latest-type ultracool
subdwarf companion known is the d/sdM9 benchmark HD 114762B ([Fe/H]=
-0.7); atmospheric models do a reasonably good job of reproducing the
medium-resolution ($\lambda$/$\Delta \lambda$ $\sim$ 3800)
near-infrared spectral features of this object, but fits to the low
resolution ($\lambda$/$\Delta \lambda$ $\sim$ 150) near-infrared
spectrum are unreliable \citep{bowler09}.

\begin{figure}
 \plotone[width=0.9\textwidth]{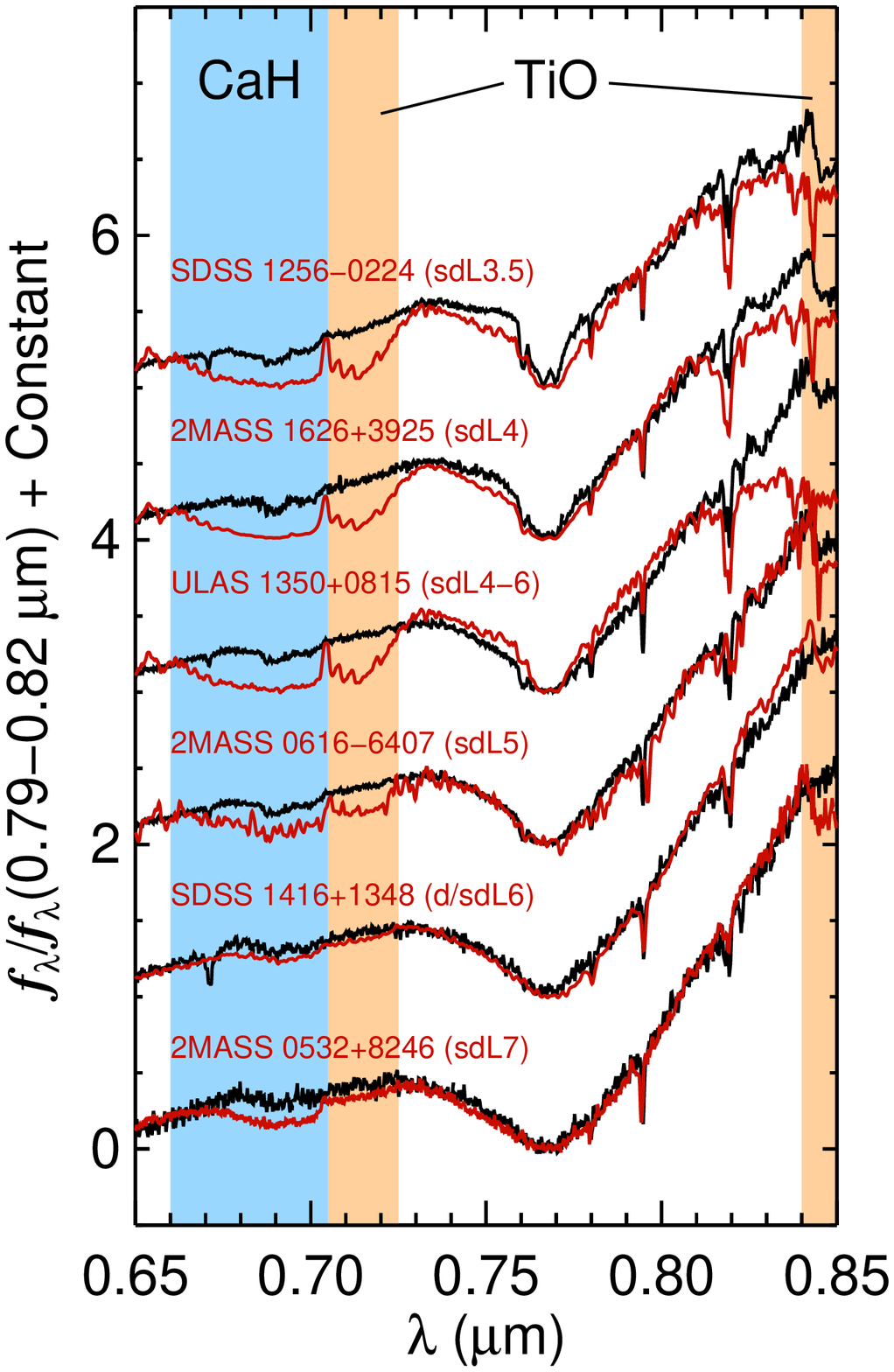}
 \caption{Optical spectra of L subdwarfs (red).  The most notable
   differences in metal-poor L dwarfs compared to ordinary field
   objects are a stronger CaH absorption band at 6800~\AA \ and
   stronger TiO absorption bands at 7100~\AA \ and 8400~\AA.  From top
   to bottom the optical spectra originate from
   \citet{2009ApJ...697..148B}, \citet{burgasser07}, \citet{Lodieu10},
   \citet{cushing09}, \citet{bowler10}, and \citet{burgasser03}.
   Comparison spectra (black) are from \citet{Kirkpatrick99}; from top
   to bottom they are 2MASS 1146+2230 (L3), 2MASS 1155+2307 (L4),
   DENIS-P J1228.2--1547 (L5), DENIS-P J1228.2--1547 (L5), 2MASS
   0850+1057 (L6), and DENIS-P J0205.4--1159 (L7).  The spectra are
   normalized between 7900~\AA \ and 8200~\AA \ and are offset by a
   constant.  \label{fig:sdL_fig} }
\end{figure}

Although atmospheric models are not yet grounded by brown dwarfs with
known metallicities, trends in the models have provided qualitative
indications of deviations from solar metallicity for a growing number
of L and T dwarfs with peculiar spectra.  The optical spectra of
peculiar L dwarfs are marked most notably by enhanced metal-hydride
and metal-oxide bands compared to normal L dwarfs of the same spectral
type (Figure \ref{fig:sdL_fig}).  These variations are likely caused
by subsolar metallicities and possibly suppressed condensate
formation. A reduced metallicity also increases collision-induced
absorption by H$_2$ (CIA H$_2$), resulting in bluer NIR
colors for a given optical spectral type.  Cloud properties also
influence the NIR colors of L dwarfs and there is no clear
way to distinguish clouds from a mild deviation from solar metallicity
from NIR colors or spectra alone \citep{burgasser08}.  Among
the $\sim$20 known objects that make up this class of ``blue L
dwarfs'' \citep{kirkpatrick10}, which is distinct from L subdwarfs,
two benchmark blue L dwarfs provide important clues about the nature
of the spectral peculiarities.  The blue L dwarf
2MASS~J17114559+4028578 (L4.5) orbits a solar-metallicity star
\citep{radigan08} and the blue L dwarf SDSS~J141624.08+134826.7
\citep[d/sdL6,][]{bowler10,schmidt_blue} has a peculiar T7.5 companion
with spectral features indicative of being mildly metal-poor
\citep{burningham10,burgasser10}; this is the first evidence that blue
L dwarfs may span a range of metallicities.  For T dwarfs, gravity and
metallicity both affect the $K$-band flux by influencing CIA H$_2$
\citep[e.g.,][]{Liu07}.  Metallicity (and to a lesser extent
gravity) also affects the $Y$-band flux, offering a way to distinguish
between these parameters \citep[e.g.,][]{leggett07}.  Ongoing
sensitive all-sky surveys like WISE and Pan-STARRS are expected to
greatly increase the census of non-solar metallicity isolated and
benchmark L and T dwarfs, enabling rigorous testing of atmospheric
models and an empirical calibration of spectral classification
schemes.

\begin{figure}
 \plotone[width=0.8\textwidth]{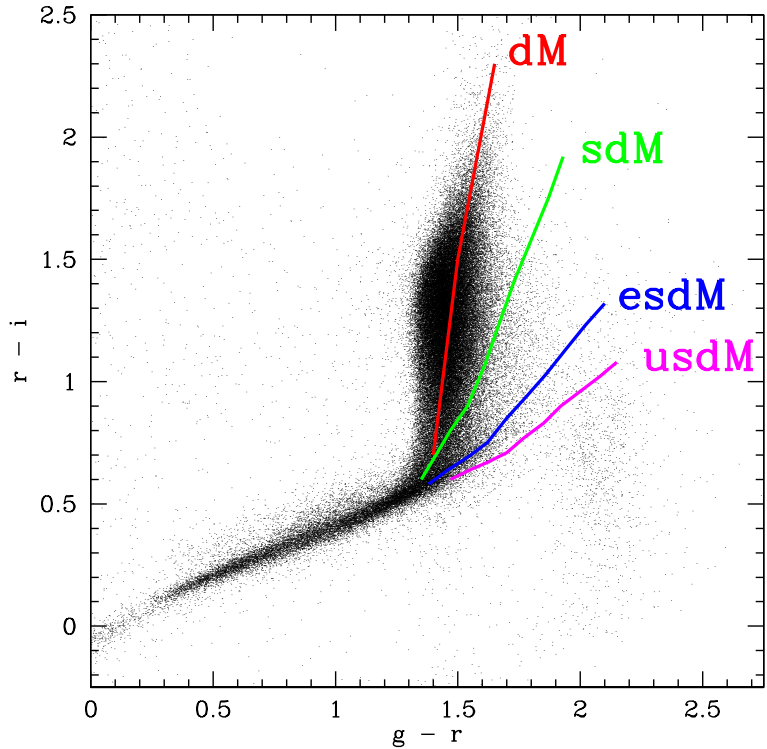}
 \caption{The $r-i$ vs. $g-r$ color-color digram for stars in the SDSS
   spectroscopic catalog. The thick colored lines show the mean loci
   for the 4 metallicity classes of M dwarfs (dM:red; sdM:green;
   esdM:blue; and usdM :purple).}
 \label{lepine}
\end{figure}

\section{The Colors and sub-Classes of Subdwarfs in SDSS}

Low--mass stars with very low metallicities, typical of the Galactic
thick disk and halo population, have a spectral energy distribution
that is significantly different from the more metal-rich disk
stars. The reason lies in the reduced absorption from metal oxide
bands, in particular TiO.  M dwarfs are classified in four
so-called ``metallicity classes'' based on the relative strengths of
their TiO bands: from the metal-rich dwarf M dwarfs (dM), to subdwarfs
(sdM), extreme subdwarfs (esdM), and the very metal-poor
ultrasubdwarfs (usdM). The classification follows the system of
\citet{1997AJ....113..806G} recently upgraded by \citet{lepine07}. The
sequence usdM$\rightarrow$esdM$\rightarrow$sdM$\rightarrow$dM is
believed to form a sequence of increasing metallicity
\citep{gizisreid97, woolf09}, with [Fe/H]$\approx$-0.5 for sdM,
[Fe/H]$\approx$-1.0 for esdM, and [Fe/H]$\lesssim$-1.5 for usdM,
although the metallicity calibration remain relatively uncertain to
this date.

\begin{figure}
 \plotone[angle=270,width=0.7\textwidth]{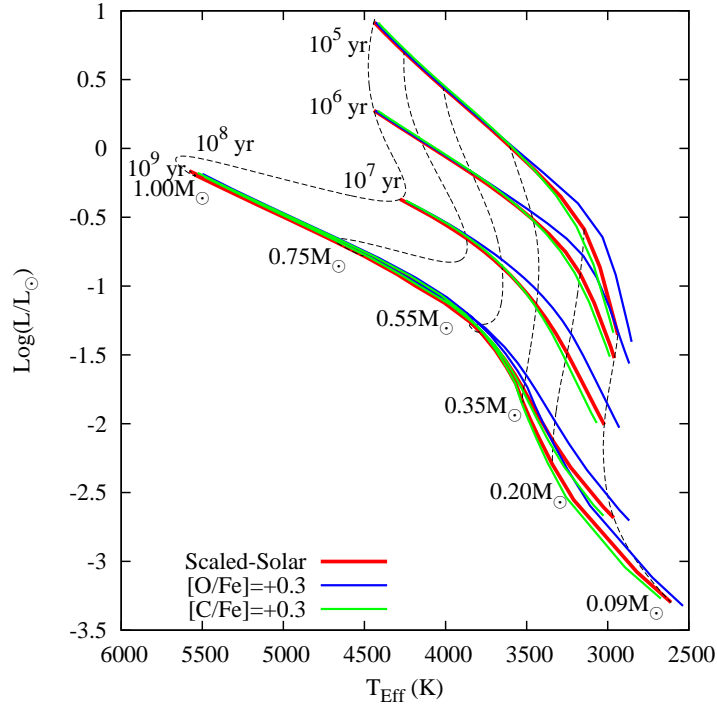}
 \caption{Luminosity vs. temperature diagram for a series of isochrones
   with masses between 0.09 and 1 $M_{\odot}$ and ages between $10^5$
   and $10^9$ years with [Fe/H]=0.  The red lines have Solar C and O
   abundances, while the blue and green lines are enhanced in O and C
   respectively.}
 \label{dotter}
\end{figure}

A recent search of the Sloan Digital Sky Survey (SDSS) spectroscopic database
(L\'epine et al. in preparation) has produced over 7,600 M
subdwarfs. Their color distribution reveals significant differences
with the metallicity class. Figure \ref{lepine} shows the $r-i$ color as a
function of $g-r$. The dots show a typical distribution for nearby
field stars, displaying the well-known ``elbow'' with a strong inflection
point at $g-r=1.4$, $r-i=0.6$. The thick colored lines show the mean
loci for the dM (red), sdM (green), esdM (blue), and usdM
(purple). There is a clear segregation as a function of the
metallicity class, which reflects the strong effect that the TiO bands
have on the spectral energy distribution. Ultrasubdwarfs simply extend
the linear relationship between $g-r$ and $r-i$, as one would expect
from a blackbody. As the metallicity increases, the "elbow" becomes
increasingly pronounced. This happens because the $r-$band gets
increasingly depressed in the more metal rich stars, as the TiO
opacity increases. This strong dependence of color on metallicity
opens the possibility of estimating metallicities in low--mass stars
based on broadband photometry alone. As it turns out, even dM show a
significant scatter in $g-r$ which could be entirely explained by
differences in metallicity. Should this be confirmed, this would
provide a formidable tool for quick and easy metallicity estimates of
low--mass stars. A proper calibration of the $g-r$ and $r-i$ color
terms as a function of metallicity should be a priority.

\section{Metallicity and Stellar Evolution Models}

``Metallicity'' loosely describes the heavy element content of a star
or stellar population.  Metallicity and [Fe/H] are often used
interchangeably, with the implicit assumption that the other heavy
elements scale with Fe as they do in the Sun.  If they don't, then
it's important to understand how changing a given element alters the
spectrum, hence the opacity, hence the effective temperature scale of
the star \citep{dotter07}.

After H and He, the two most abundant elements in the sun by mass or
number fraction are C and O \citep[e.g.,][]{asplund09}. When C or O
is enhanced relative to solar at fixed [Fe/H] the most dramatic effect
appears in the molecular opacities. Figure \ref{dotter} shows a series of
isochrones with masses between 0.09 and 1 $M_{\odot}$ and ages between
$10^5$ and $10^9$ years with [Fe/H]=0. As Figure \ref{dotter} indicates, on the
one hand, enhancing C actually makes the lowest mass stars hotter
while, on the other hand, enhancing O makes them cooler. This behavior
can be understood in terms of the contribution of water molecules to
the opacity.  When analyzing the physical properties of low mass stars
with effective temperatures below about 4,000K it is important to
consider that non-solar abundance ratios can skew the results.

\section{Metallicity and Atmosphere Models}

One of the primary tools for measuring metallicities of low--mass
objects is the comparison of data to synthetic spectra (created using
atmospheric models). Calculating such models is common practice
\citep[e.g.,][]{2010Ap&SS.328..331S} using modern atmosphere codes
\citep{1999JCoAM.109...41H}.  For low--mass objects in general, it is
crucial to account for dust formation in the atmosphere. This dust
formation needs to be treated as a microphysical growth and
destruction process \citep[e.g.][]{2008A&A...485..547H}.  Furthermore,
molecules, both as an opacity source and as material that affects the
equation of state, need to be accounted for with accurate input data
such as formation constants or related quantities and line lists or
equivalent opacity data.  The situation is further complicated when
calculating models specifically for low metallicity objects, such as
the models of \citet{2009A&A...506.1367W}. The decreasing metal
content does not change or even simplify any of the main physical
processes, but, in contrast, adds another dimension of parameter
space. In particular, dust keeps forming in significant amounts down
to metallicities of about [Fe/H]=-4.0 \citep{2009A&A...506.1367W}.

\begin{figure}
 \plotone[width=0.8\textwidth]{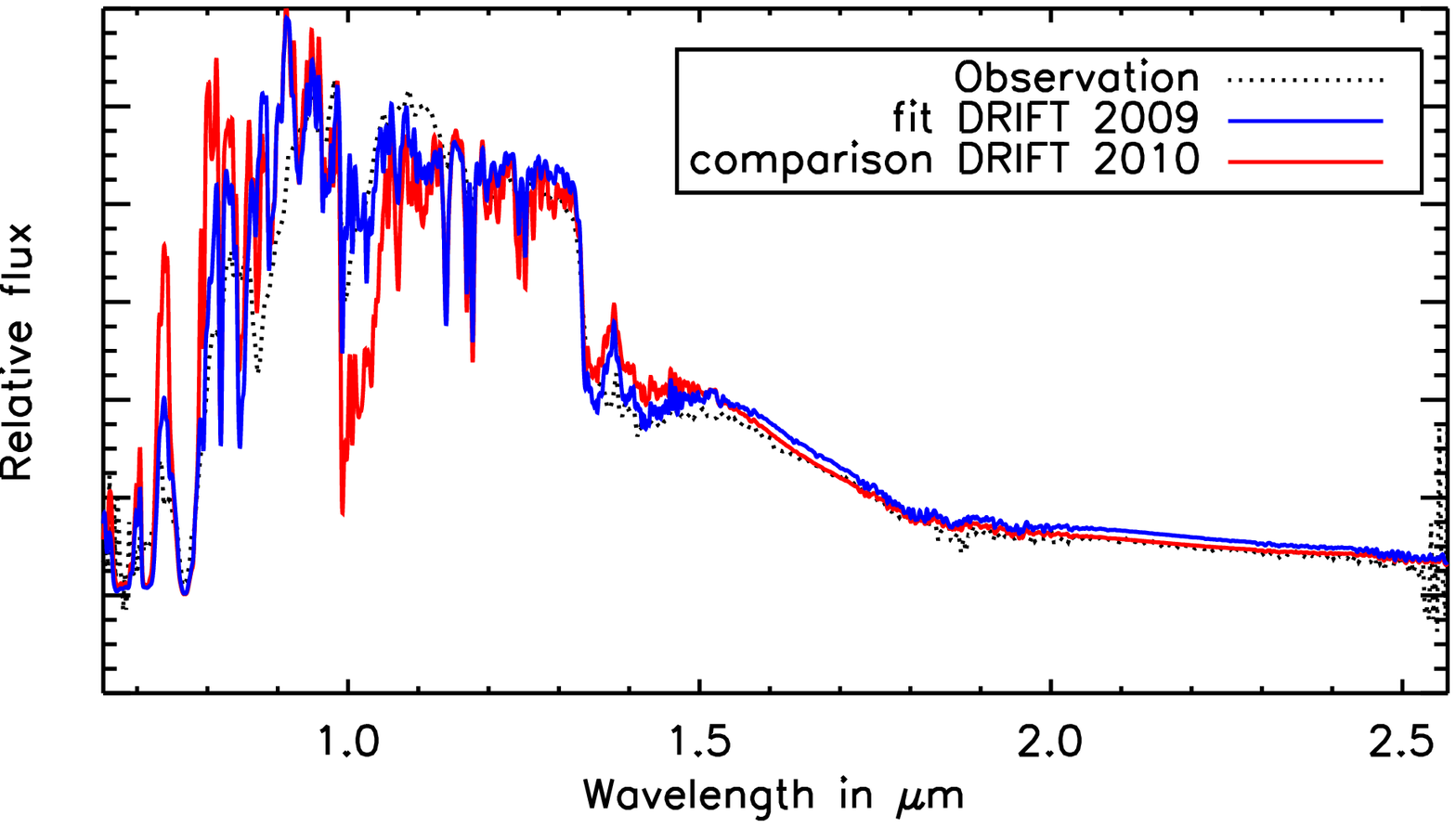}
 \caption{The sdL4 dwarf 2MASS1626+3925 \citep[black, dotted;][]{2004ApJ...614L..73B}
and a fit with the DRIFT 2009 models \citep[blue, solid; ][]{2009A&A...506.1367W}.
The comparison with the DRIFT 2010 models \citep[red, solid;][]{witte10}
is a comparison with the same model parameters showing the differences
in model details. Both models have $T_{\rm eff}=2100$K, $\log(g)$=5.0 and
a metallicity [Fe/H]$=-1.5$.}
 \label{andreas}
\end{figure}

Recently, it has become possible to apply synthetic spectra to
observations of L subdwarfs and to attempt to measure metallicities
\citep{2009ApJ...697..148B}.  However, the quality of the fits and the
derived metallicities still vary \citep{witte10} depending on the
quality of the implemented physics (see also Fig. \ref{andreas}).
However, it is worth noting that the derived metallicities of the sdL
class do not need to be the same of the sdM class as derived by
e.g. \citet{1997AJ....113..806G} or \citet{1999PhDT........12S}.
Measuring metallicities directly at a resolution of 0.1 dex or higher
has not been attempted yet since the molecular background lines add
too much uncertainty.

\section{Conclusions}
During the first half of the last century, spectroscopic observations
and radiative transfer theory began to unlock the composition of
stars.  Determining the metallicity of stars has been very important
to a wide range of astronomical investigations, from planetary to
cosmological scales.  Yet, despite the progress made for most of the
main sequence, measuring the metallicity of the Galaxy's most populous
members, M dwarfs, remains a daunting task.  

At the start of this century, astronomers are beginning to unlock the
metal content of these stars.  However, there is much work to be done
by both observers and theorists.
Observationally, there are promising new results suggesting that IR
observations may be important for estimating metallicities.  This is
strengthened by the relative agreement between models and spectra in
this regime.  However, these methods need further testing (with M dwarf
binaries or clusters).  In the optical bandpass, the relative
metallicity classes described in Section 5 display a clear separation in
photometric colors.  This will be crucial for estimating the metal
content of these stars in the next generation of surveys, which will
be largely photometric.  Yet, these classes have not been rigorously
tied to an absolute metallicity scale.   Once this occurs, the
chemical composition of M dwarfs will be a powerful tool for studying
the Galaxy and identifying the most likely exoplanet hosts.
Identifying new benchmarks, for both M dwarfs and brown dwarfs, will
be crucial in calibrating optical observations. 

On the theoretical front, new line lists, opacity calculations and the
inclusion of dust grains have resulted in better agreement with
observations.  The effects of carbon and oxygen abundance differences
can be modeled and explain observations of the lower main sequence of
globular clusters.  As computational power and techniques advance,
these models will grow in sophistication and should offer a more
realistic picture of the important physics within these stars.

Unlocking the metallicity of M dwarfs will profoundly benefit the astronomical
community in a variety of ways, such as identifying exoplanet hosts and
studying chemical evolution.  The work presented here represents the
first steps in solving this problem. 

\acknowledgements The authors would like to thank the Cool Stars 16
SOC for the opportunity to hold this productive splinter session. BPB
gratefully acknowledges the Cool Stars 16 Accommodation Stipend Award,
funded by the NASA Astrobiology Institute.

\end{document}